\documentclass[%
 reprint,
 amsmath,amssymb,
pra,
]{revtex4-2}

\usepackage{graphicx}
\usepackage{dcolumn}
\usepackage{bm}
\usepackage{color}
\usepackage{float}

\begin{document}
\hyphenpenalty=900
\title{Tunable Rotation-Associated Slow-to-Fast Light Conversion via Optomagnonic Coupling}

\author{Jingyu Liu\textsuperscript{1,2}}

\author{Shirong Lin\textsuperscript{1,2}}
\email{shironglin@gbu.edu.cn}

\affiliation{\textsuperscript{1}School of Physical Sciences, Great Bay University, Dongguan 523000, China}
\affiliation{\textsuperscript{2}Great Bay Institute for Advanced Study, Dongguan 523000, China}
\date{\today}

\begin{abstract}
Cavity optomechanics has enabled slow-to-fast light conversion, but traditional optomechanic systems suffer from limited tunability due to fixed mechanical frequencies. To address this constraint, we introduce a magnon degree of freedom into an optomechanical system, constructing a system that integrates photons, phonons, and magnons. We establish the theoretical model of the optomagnonic-Laguerre-Gaussian rotational system, and present numerical simulations of Fano resonances and group delay. By manipulating the magnon degree of freedom, we not only achieve slow-to-fast light conversion associated with magnons but also successfully realize such conversion effects associated with mechanical rotation—this achievement effectively overcomes the inherent tunability limitations of pure optomechanical systems and expands the frequency coverage of light conversion effects. Notably, we numerically demonstrate bidirectional light speed conversion (slow-to-fast and fast-to-slow) via continuous control field frequency modulation to tune cavity mode detuning. Additionally, our results show that adjusting optomagnonic parameters enables dynamic switching between slow light and fast light at multiple frequencies. This work provides a flexible platform for multi-frequency light speed control, with potential applications in all-optical networks and quantum communications.
\end{abstract}

\maketitle

\section{Introduction}
Optical-frequency cavity optomechanics, driven by radiation pressure, gives rise to optomechanically induced transparency (OMIT)—a quantum interference effect where a strong control field opens a narrow transparency window for a weak probe field \cite{PRA2010Electromagnetically,Sci2010Optomechanically}. Optomechanical systems extend beyond macroscopic mechanical setups, including Rydberg-atom \cite{PRL2014Optomechanical} and 2D material platforms \cite{PRB2016Light}. OMIT induces drastic phase dispersion, leading to slow light (characterized by reduced group velocity, \(\tau_g>0\)) or fast light (characterized by negative group delay, \(\tau_g<0\)) \cite{PRA2015Tunable,PRA2020Tunable}. Slow light is critical for controllable optical signal delay, underpinning optical communication buffering \cite{NP2008Slow,PRL2003Observation}, quantum state storage \cite{JPDAP2022Significantly}, dynamical Casimir effects \cite{PRA2008Optical} and high-sensitivity magnetometry \cite{OE2019Magnetically}. Fast light, meanwhile, enables signal advancement processing \cite{NP2008Slow}, high-resolution sensing \cite{LPR2010Slow}, and efficient phase regulation in microwave photonics \cite{LPR2009Slow}. The core value of slow-to-fast light conversion lies in dynamically switching group velocity via external control: this adaptability allows it to meet signal processing requirements across scenarios, serving as key technical support for all-optical networks, quantum communications, and high-precision sensing \cite{OE2005Gain,PRA2020Tunable,PRA2021Phase,PRXQ2021Quantum,JLT2023Fano,APL2023Controllable}. However, traditional optomechanical systems rely on fixed mechanical frequencies or cavity detunings, which restricts real-time light speed tuning \cite{PRA2015Tunable,JLT2023Fano}.

To overcome this limitation, cavity optomagnonic systems—integrating magnons \cite{PR1948On,NL2015Skyrmion,PRL2018Magnon,PRL2021magnonic,PRB2023Bidirectional,guo2024quantum,APL2024Magnonics,IEEE2025Reconfigurable,NC2025Coherent} (quantized spin excitations) with photons—offer superior controllability: magnon frequencies are tunable via external magnetic fields \cite{PRB2006Magnetization,PRL2011AllMagnonic,MTE2023A,PRRSPL2022Cavity}, and insulators like yttrium iron garnet (YIG) exhibit ultra-low magnon damping \cite{PRRSPL2022Cavity,PRB2016Optomagnonics}. More importantly, magnons possess inherent advantages that make them ideal for hybrid light-matter systems: they support nonreciprocal effects in cavity architectures \cite{PRB2021Nonreciprocal,PRA2022Nonreciprocal1,PRA2022Nonreciprocal2}, enable the construction of a parity-time symmetric system \cite{PRA2019Enhanced} and facilitate wideband conversion via nonlinearity in hybrid quantum setups \cite{APL2023Controllable,NPJS2024Wideband}. Most existing optomagnonic studies focus on resonant magnon-photon coupling in the microwave regime (frequency matching), with key breakthroughs including strong coupling in YIG-microwave cavity hybrids \cite{PRL2010Strong,PRB2010Size,PRL2013High,PRL2014Hybridizing,APE2019Hybrid} and magnon-polariton formation \cite{JAP2020Universal,PR2022Quantum}. For telecommunication-relevant optical frequencies, however, coupling is non-resonant (large frequency mismatch), which requires parametric interactions mediated by the magneto-optical effect \cite{PRA2016Coupled,PRL2016Cavity,PRB2016Optomagnonics,NP2019Alloptical,LPR2025Microwave}.

Building on Laguerre-Gaussian (LG) rotational cavities—an optomechanical system with a fixed mechanical frequency where the orbital angular momentum (OAM) of LG beams modulates optorotational coupling for light speed control \cite{JLT2023Fano}—we introduce a highly  tunable optomagnonic coupling. By adjusting optomagnonic parameters, we achieve slow-to-fast light conversion not only at the magnon frequency but also at the frequency of mechanical rotation. Notably, we demonstrate bidirectional light speed conversion (both slow-to-fast and fast-to-slow conversions) by continuously modulating the control field frequency to adjust its cavity mode detuning. This extends magnetostrictive slow light \cite{OE2019Magnetically} and leverages non-resonant parametric coupling to overcome the tunability limitations of pure optomechanics.

This paper is structured as follows: Section II establishes the theoretical model of the optomagnonic-LG rotational system, deriving the Hamiltonian and Heisenberg-Langevin equations that integrate photons, phonons, and magnons. Section III presents numerical simulations of Fano resonances, group delay, and slow-to-fast light conversion. Conclusions are drawn in Section IV.
\section{Model and Methods}
\subsection{Model}
Our system consists of an LG rotational cavity and a YIG sphere (hosting magnons) inside it, as shown in Fig.1 . The cavity comprises two spiral phase elements: a partially transmissive input coupler (IC) on the left and a fully reflective rear mirror (RM) on the right. The RM can rotate about the z axisis. The IC retains the topological charge, the orbital angular momentum quantum number of the LG beam denoted as \(l\), of transmitted light but removes a fixed \(2l\) from reflected light; in contrast, the RM adds a fixed \(2l\) to reflected light—this modulation ensures stable LG mode establishment. Notably, fabricated spiral phase plates can generate LG beams with topological charges up to 1000 \cite{JO2013Generation}. The optorotational coupling Hamiltonian is expressed as: \(H_{o\phi}=-\hbar g_{\phi}a^{\dagger}a\phi\), where \(a^\dagger\) and \(a\) are the photon creation and annihilation operators; \(\phi\) is the angular displacement of the RM around the cavity axis; \(g_{\phi}=\frac{cl}{L}\sqrt{\frac{\hbar}{I\omega_{\phi}}}\) denotes the optorotational coupling strength, depending on the speed of light \(c\), topological charge \(l\), cavity length \(L\), RM moment of inertia \(I\) (with \(r\) the RM radius and \(m\) the RM mass), and rotational frequency \(\omega_\phi\) the latter corresponding to an energy of \(\hbar\omega_\phi\) for a single rotational phonon of the mechanical mode \cite{JPB2021Entanglement,JLT2023Fano}. The optorotational interaction originates from torque exerted by the OAM of LG beams on the RM, enabling coherent coupling between optical and rotational mechanical modes \cite{PRL2007Using,JLT2023Fano}.

\begin{figure}[htbp!]
    \centering
    \includegraphics[width=0.45\textwidth]{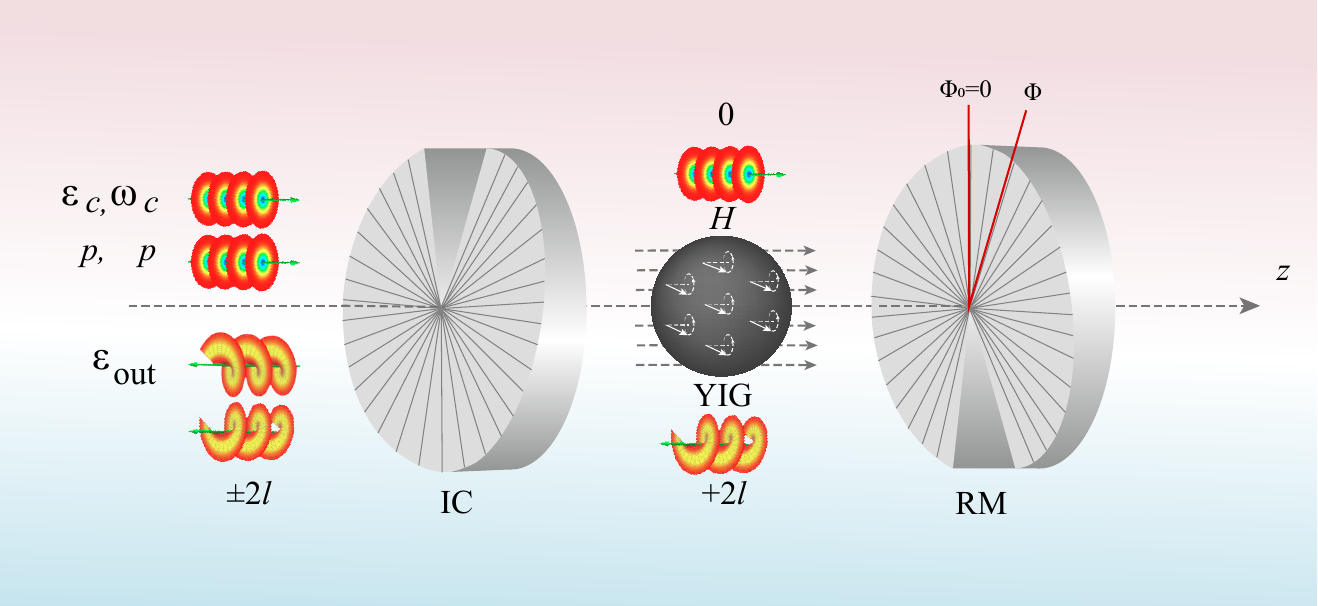}
    \caption{Schematic of the optorotational-optomagnonic hybrid model. The incident light is a circularly polarized Gaussian beam: the beam’s orbital angular momentum is transferred to the RM’s rotational angular momentum.}
    \label{model}
\end{figure}

The magnon frequency is governed by the external bias magnetic field $H$ and gyromagnetic ratio \(\gamma\), with the quantitative relationship \(\omega_m = \gamma H\)—corresponding to the ferromagnetic resonance frequency of the Kittel mode, a collective lowest-order magnon excitation \cite{PR1948On}. Non-resonant optomagnetic coupling satisfies both energy and angular momentum conservation, which is achieved by coupling a single circularly polarized light mode with a YIG microsphere \cite{PRA2016Coupled}; the detailed derivation is presented in the subsequent sections.  Our system incorporates three degrees of freedom: cavity mode, rotation, and magnons. Additionally, we apply a strong control field (\(i\hbar \varepsilon_c a^\dagger e^{-i\omega_c t} - i\hbar \varepsilon_c a e^{i\omega_c t}\)) and a weak probe field (\(i\hbar \varepsilon_p a^\dagger e^{-i\omega_p t} - i\hbar \varepsilon_p a e^{i\omega_p t}\)) to the system, where \(\varepsilon_c = \sqrt{2\kappa_a P_c/\hbar\omega_c}\) and \(\varepsilon_p = \sqrt{2\kappa_a P_p/\hbar\omega_p}\) are the control and probe field amplitudes, \(\kappa_a\) is the cavity photon decay rate, and \(P_c(P_p)\) is the control (probe) field power. The total Hamiltonian \(H\) is:
\begin{align}
H=&\hbar\omega_a a^{\dagger}a + \frac{1}{2}\hbar\omega_\phi(L_{z}^{2}+\phi^{2})-\hbar g_{\phi}a^{\dagger}a\phi \notag\\
&+\hbar\omega_m m^{\dagger}m+\hbar g_{m}a^{\dagger}a(m + m^{\dagger}) \notag\\
&+i\hbar\varepsilon_{c}(a^{\dagger}e^{-i\omega_c t}-a e^{i\omega_c t})+i\hbar \varepsilon_p (a^\dagger e^{-i\omega_p t}- a e^{i\omega_p t}),
\end{align}
where the terms on the right-hand side represent: the free Hamiltonian of the cavity photon mode, the free Hamiltonian of the RM’s rotational mode, optorotational coupling between cavity photons and the RM, the free Hamiltonian of magnons and optomagnonic coupling, and coupling between the cavity mode and the control/probe fields. Specifically: \(\omega_a\) is the cavity mode frequency, \(\omega_m\) the magnon frequency; \(g_m\) is the optomagnonic coupling strengths; \(a(a^\dagger)\) and \(m(m^\dagger)\) are the photon and magnon annihilation (creation) operators; \(L_z/\phi\) are the RM’s dimensionless angular momentum/angular displacement, satisfying the commutation relation \([\phi, L_z] = i\).
\subsection{Methods}
In the rotating frame of the control field frequency (via the unitary transformation \(U = e^{-i\omega_c(a^\dagger a)t}\)), the Hamiltonian becomes:
\begin{align}
H=&\hbar\Delta_aa^{\dagger}a + \frac{1}{2}\hbar\omega_\phi(L_{z}^{2}+\phi^{2})-\hbar g_{\phi}a^{\dagger}a\phi \notag\\
&+\hbar\omega_m m^{\dagger}m+\hbar g_{m}a^{\dagger}a (m^{\dagger}+m) \notag\\
&+i\hbar\varepsilon_{c}(a^{\dagger}-a)+i\hbar\varepsilon_{p}(a^{\dagger}e^{-i\delta t}-ae^{i\delta t}),
\end{align}
where \(\Delta_a = \omega_a - \omega_c\) (cavity detuning from the control field) and \(\delta = \omega_p - \omega_c\) (probe detuning from the control field). Considering cavity losses, magnon damping, and intrinsic damping of the RM’s rotation, the quantum Langevin equations for the operators in the Heisenberg picture are:
\begin{align}
\frac{d}{d t}a=&-\left(i\Delta_{a}+\kappa_{a}\right)a+i g_{\phi}a\phi-ig_{m}a(m^\dagger+m)\notag\\
               &+\varepsilon_{c}+\varepsilon_p e^{-i\delta t}+\sqrt{2\kappa_a}a^{in},\notag\\
\frac{d}{d t}L_{z}=&-\omega_{\phi}\phi+ g_{\phi}a^{\dagger}a-\kappa_{\phi}L_{z}+\eta^{in},\notag\\
\frac{d}{d t}\phi=&\omega_\phi L_{z},\notag\\
\frac{d}{d t}m=&-(i\omega_{m}+\kappa_m)m-ig_{m}a^{\dagger}a+\sqrt{2\kappa_m}m^{in},
\end{align}
where \(\kappa_a\) and \(\kappa_m\) are the cavity photon and magnon decay rates; \(\kappa_\phi\) is the intrinsic damping rate of the rotational mode; \(a^{in}\), \(m^{in}\), and \(\eta^{in}\) are the input noise operators for the cavity mode, magnons, and rotation, respectively (all with zero mean).

Under the assumption that the control field is stronger than the probe field, the solutions to the equations can be expressed as the sum of a steady-state average and a small fluctuation term: \(O = O_s + \delta O\) (where \(O = a,L_z,\phi,m\)). Substituting this into the quantum Langevin equations and neglecting high-order small terms (e.g., \(\delta O_1\delta O_2\)), we obtain equations for the steady-state averages and linearized quantum Langevin equations. The steady-state average equations are:
\begin{align}
L_{z s}&=0,\notag\\
\phi_{s}&=\frac{g_{\phi}|a_{s}|^{2}}{\omega_{\phi}},\notag\\
m_s&=\frac{-ig_m|a_s|^2}{i\omega_m+\kappa_m}, \notag\\
\quad a_{s}&=\frac{\varepsilon_{c}}{\kappa_{a}+i\Delta},
\end{align}
where \(\Delta = \Delta_a - g_\phi \phi_s + g_m(m_s^* + m_s)\) denotes the effective detuning, and $m_s^*$ represents the complex conjugate of $m_s$. The linearized equations retaining only first-order terms of \(\delta O\) are:
\begin{align}
\frac{d\delta a}{d t}=&-\left(i\Delta + \kappa_{a}\right)\delta a+i g_{\phi}a_s\delta\phi-ig_{m}a_s(\delta m^\dagger+\delta m)\notag\\
                      &+\varepsilon_p e^{-i\delta t}+\sqrt{2\kappa_a}a^{in},\notag\\
\frac{d\delta L_z}{d t}=&-\omega_{\phi}\delta\phi+ g_{\phi}(a_s^\dagger\delta a + a_s\delta a^\dagger)-\kappa_{\phi}\delta L_{z}+\eta^{in},\notag\\
\frac{d\delta \phi}{d t}=&\ \omega_\phi \delta L_{z},\notag\\
\frac{d\delta m}{d t}=&-(i\omega_{m}+\kappa_m)\delta m-ig_{m}(a_s^\dagger\delta a + a_s\delta a^\dagger)\notag\\
                      &+\sqrt{2\kappa_m}m^{in}.
\end{align}

Since the control and probe fields are weak but classical coherent fields, we identify operators with their expectation values (\(\delta O(t) \equiv \langle\delta O(t)\rangle\)) and drop quantum/thermal noise terms (which average to zero). We adopt the following ansatz for the fluctuations:
\begin{align}
\delta a(t)=&A^{-}e^{-i\delta t}+A^{+}e^{i\delta t},\notag\\
\delta a^{*}(t)=&(A^{-})^*e^{i\delta t}+(A^{+})^*e^{-i\delta t},\notag\\
\delta m(t)=&M^{-}e^{-i\delta t}+M^{+}e^{i\delta t},\notag\\
\delta m^{*}(t)=&(M^{-})^*e^{i\delta t}+(M^{+})^*e^{-i\delta t},\notag\\
\delta \phi(t)=&\Phi e^{-i\delta t}+\Phi^{*}e^{i\delta t}.
\end{align}
From this ansatz, the transmittance of the probe field at frequency \(\omega_c + \delta\) depends only on \(A^-\) (i.e., the anti-Stokes field), while the Stokes field \(A^+\) can be analyzed using the same approach \cite{Sci2010Optomechanically,OE2019Magnetically}. Solving for \(A^-\) yields:
\begin{equation}
A^{-}=\frac{\varepsilon_{p}(\kappa_{a}-i\delta -i\Delta+F)}{F^2+(\kappa_a-i\delta)^2-(\Delta -iF)^2},
\end{equation}
where \(F=\frac{ig_\phi^2a_s^2\omega_\phi}{\omega_\phi^2-\delta^2-i\kappa_\phi\delta}+\frac{ig_m^2a_s^2\omega_m}{\omega_m^2-(\delta -i\kappa_m)^2}\). According to the input-output relation \cite{Book2004Quantum}, the amplitude of the output probe field is \(\varepsilon_{out}=2\kappa_{a}A^{-}/\varepsilon_{p}\), where the real and imaginary parts of \(\varepsilon_{out}\) represent the probe field’s absorption and phase dispersion, respectively. Additionally, the probe field’s transmission coefficient is defined as \(t_{p}=1 - 2\kappa_{a}A^{-}/\varepsilon_{p}\), with phase \(\phi_t = \arg(t_p)\). The group delay—calculated by taking the partial derivative of the phase with respect to the probe frequency—is:
\begin{equation}
\tau_g=\frac{\partial\phi_t(\omega_p)}{\partial\omega_p},
\end{equation}
where \(\tau_g>0\) indicates slow light and \(\tau_g<0\) indicates fast light.

\subsection{Optomagnonic Coupling}
\subsubsection{Coupling Mechanism via Magneto-Optical Effect}
Non-resonant coupling between magnons and optical photons (with \(\omega_{\text{opt}} \gg \text{magnon frequency } \Omega_m\)) arises from the magneto-optical effect, which is described by the magnetization-dependent dielectric tensor \(\overleftrightarrow{\varepsilon}(\mathbf{M})\) \cite{PRA2016Coupled,PRB2016Optomagnonics}. For a cubic magnetic insulator (e.g., YIG) with equilibrium magnetization \(\mathbf{M}_0 \parallel \hat{z}\), the linear-order correction to the dielectric tensor—dominant for non-resonant coupling—is given by:
\begin{equation}
\overleftrightarrow{\varepsilon}_1(\mathbf{M}) = \begin{pmatrix}
0 & -ifM_s & ifM_y \\
ifM_s & 0 & -ifM_x \\
-ifM_y & ifM_x & 0
\end{pmatrix},
\end{equation}
where \(M_s = |\mathbf{M}_0|\) (saturation magnetization), \(f\) is a material-dependent magneto-optical constant, and \(M_x, M_y\) are the dynamical (fluctuating) components of magnetization associated with magnons.

The interaction Hamiltonian is derived from the electromagnetic energy density \(\mathcal{H} = \frac{1}{2}\mathbf{E} \cdot \mathbf{D}\) (where \(\mathbf{D} = \varepsilon_0 \overleftrightarrow{\varepsilon}(\mathbf{M}) \cdot \mathbf{E}\)), with the dominant contribution coming from Brillouin Light Scattering \cite{JLT2023Fano,PRA2016Coupled}:
\begin{equation}
\hat{H}_{\text{BLS}} = \hbar \sum_{pq\eta} \hat{a}_p^\dagger \hat{a}_q \left( G_{pq\eta}^+ \hat{m}_\eta + G_{pq\eta}^- \hat{m}_\eta^\dagger \right) + \text{h.c.},
\end{equation}
where \(\hat{a}_p^\dagger/\hat{a}_q\) are photon creation/annihilation operators for optical modes \(p/q\); \(\hat{m}_\eta^\dagger/\hat{m}_\eta\) are magnon creation/annihilation operators; and \(G_{pq\eta}^\pm\) are coupling coefficients that encode mode overlap and material properties \cite{PRRSPL2022Cavity},
\begin{equation}
G_{pq\eta}^\pm = \frac{\mathcal{G} M_s}{2} \int_V d^3\mathbf{r}\, w_\eta^{(*)}(\mathbf{r}) \cdot \left[ u_p^*(\mathbf{r}) \times u_q(\mathbf{r}) \right],
\end{equation}
where \(u_p(\mathbf{r})/u_q(\mathbf{r})\) are optical mode functions, \(w_\eta(\mathbf{r})\) is the magnon mode function (uniform for the Kittel mode, with \(w_\eta(\mathbf{r}) \parallel \hat{z}\) to align with \(\mathbf{M}_0\)), and \(\mathcal{G}\) is a material-dependent constant. Non-zero coupling occurs only if the cross product \(u_p^* \times u_q\) has a non-vanishing component along \(w_\eta(\mathbf{r})\).
\subsubsection{Coupling for TE/TM Modes}
Optical cavities typically support transverse electric (TE) and transverse magnetic (TM) modes as linear polarization eigenmodes \cite{PRRSPL2022Cavity}. For propagation along \(\hat{z}\) (matching \(\mathbf{M}_0 \parallel \hat{z}\)): The electric field of TE mode \(\mathbf{E}_{\text{TE}}\) is perpendicular to the propagation plane (e.g., \(u_{\text{TE}}(\mathbf{r}) = \hat{e}_x e^{ik_z z}\), normalized to the optical mode volume \(V_{\text{opt}}\)); The electric field of TM mode \(\mathbf{E}_{\text{TM}}\) lies within the propagation plane (e.g., \(u_{\text{TM}}(\mathbf{r}) = \hat{e}_y e^{ik_z z}\)). The coupling coefficient \(G_{pq\eta}^\pm\) depends on the integral of \(u_p^* \times u_q\) along \(\hat{z}\): For same-mode combinations (TE-TE, TM-TM), the cross product vanishes, leading to \(G_{\text{TE,TE},\eta}^\pm = G_{\text{TM,TM},\eta}^\pm = 0\); For cross-mode combinations (TE-TM), the cross product yields a non-vanishing \(\hat{z}\)-component, resulting in \(G_{\text{TE,TM},\eta}^\pm \neq 0\). This implies TE/TM modes require polarization conversion (\(\text{TE}\longleftrightarrow \text{TM}\)) to achieve non-zero optomagnonic coupling.
\subsubsection{Coupling for Circularly Polarized Light (CPL)}
CPL is an eigenstate of photon angular momentum (with spin \(\pm\hbar\)) and arises from phase-shifted superpositions of TE and TM modes \cite{PRB2016Optomagnonics,OE2019Magnetically}. For propagation along \(\hat{z}\), the two CPL modes are: left-circular polarization (LCP, \(\sigma=+\)): \(u_+ (\mathbf{r}) = \frac{\hat{e}_x - i\hat{e}_y}{\sqrt{2}} e^{ik_z z}\) (spin angular momentum \(+1\) along \(\hat{z}\)); right-circular polarization (RCP, \(\sigma=-\)): \(u_- (\mathbf{r}) = \frac{\hat{e}_x + i\hat{e}_y}{\sqrt{2}} e^{ik_z z}\) (spin angular momentum \(-1\) along \(\hat{z}\)). Unlike TE/TM modes, CPL modes exhibit non-zero coupling even for \(p=q\) (same CPL mode). For LCP (\(\sigma=+\)), phase factors cancel, and the cross product simplifies to \(\hat{e}_z\), leading to:
\begin{equation}
G_{+,+,\eta}^\pm = \frac{\mathcal{G} M_s w_0 V_{\text{opt}}}{4} \neq 0.
\end{equation}
For RCP (\(\sigma=-\)), the phase factors similarly cancel, while the cross product simplifies to \(-\hat{e}_z\), resulting in:
\begin{equation}
G_{-,-,\eta}^\pm = -\frac{\mathcal{G} M_s w_0 V_{\text{opt}}}{4} \neq 0.
\end{equation}
This non-zero coupling arises because CPL, as an angular momentum eigenstate, naturally matches the spin-1 nature of magnons \cite{OE2019Magnetically}. No polarization conversion is needed for CPL-magnon coupling—the same CPL mode suffices to generate interaction. In this work, the optomagnonic coupling strength is uniformly denoted as \(g_m\). The optomagnonic coupling term in the Hamiltonian is the core mechanism for the magnetically controllable light manipulation in this work. This light-magnetism coupling is intrinsically bidirectional: while we focus on tailoring optical fields via spin dynamics here, it also enables all-optical control of magnetic order and chiral spin textures, as demonstrated in prior works \cite{AM2024Single,zhang2026skyrmion}.
\subsection{Linear Stability Analysis}
The slow-to-fast light conversion effect in our system fundamentally relies on the stability of the system’s steady state. For a dynamical system subjected to tiny perturbations, it is defined as linearly stable if the perturbations decay exponentially with time and the system returns to its steady state, where the optical response and group delay can be stably observed in experiments. Conversely, the system is unstable if the perturbations grow exponentially and the steady state collapses. We base our stability analysis on the linearized quantum Langevin equations describing the system’s fluctuation dynamics, where zero-mean quantum and thermal noise terms are neglected (as they do not influence the system’s intrinsic stability). These equations are explicitly written as:
\begin{equation}
\begin{aligned}
\frac{\mathrm{d}\delta a}{\mathrm{d}t} &= -(i\Delta + \kappa_a)\delta a + ig_\phi a_s\delta\phi - ig_m a_s(\delta m + \delta m^*), \\
\frac{\mathrm{d}\delta a^*}{\mathrm{d}t} &= (i\Delta - \kappa_a)\delta a^* - ig_\phi a_s^*\delta\phi + ig_m a_s^*(\delta m + \delta m^*), \\
\frac{\mathrm{d}\delta L_z}{\mathrm{d}t} &= g_\phi(a_s^*\delta a + a_s\delta a^*) - \kappa_\phi\delta L_z - \omega_\phi\delta\phi, \\
\frac{\mathrm{d}\delta\phi}{\mathrm{d}t} &= \omega_\phi\delta L_z, \\
\frac{\mathrm{d}\delta m}{\mathrm{d}t} &= -ig_m(a_s^*\delta a + a_s\delta a^*) - (i\omega_m + \kappa_m)\delta m, \\
\frac{\mathrm{d}\delta m^*}{\mathrm{d}t} &= ig_m(a_s\delta a + a_s^*\delta a^*) + (i\omega_m - \kappa_m)\delta m^*.
\end{aligned}
\end{equation}

Building on this set of linearized equations, we define the perturbation vector of the system as:
\begin{equation}
\boldsymbol{u}(t) = \left[\delta a,\ \delta a^*,\ \delta L_z,\ \delta \phi,\ \delta m,\ \delta m^*\right]^T.
\end{equation}
With this definition, the linearized dynamical equations can be recast into the standard matrix form of a first-order linear homogeneous ordinary differential equation, which is the universal formalism for stability analysis in cavity optomechanics and magnonics:
\begin{equation}
\dot{\boldsymbol{u}}(t) = \boldsymbol{M} \cdot \boldsymbol{u}(t).
\end{equation}
Here $\boldsymbol{M}$ is the 6×6 dynamical evolution matrix of the system, which fully encodes the coupling, dissipation, and detuning dynamics of the cavity photon, rotational phonon, and magnon modes, with the explicit form:
\begin{widetext}
\begin{equation}
\boldsymbol{M} = \begin{pmatrix}
-(i\Delta+\kappa_a) & 0 & 0 & ig_\phi a_s & -ig_m a_s & -ig_m a_s \\
0 & (i\Delta-\kappa_a) & 0 & -ig_\phi a_s^* & ig_m a_s^* & ig_m a_s^* \\
g_\phi a_s^* & g_\phi a_s & -\kappa_\phi & -\omega_\phi & 0 & 0 \\
0 & 0 & \omega_\phi & 0 & 0 & 0 \\
-ig_m a_s^* & -ig_m a_s & 0 & 0 & -(i\omega_m+\kappa_m) & 0 \\
ig_m a_s & ig_m a_s^* & 0 & 0 & 0 & (i\omega_m-\kappa_m)
\end{pmatrix}.
\end{equation}
\end{widetext}

This matrix formalism is fully consistent with the established stability analysis framework for cavity magnomechanical systems in previous magnetostrictive slow light studies [1], where the stability of the hybrid system is also determined by the eigenvalue properties of the linearized dynamical matrix.

With the dynamical matrix established, we define the rigorous stability criterion of the system in accordance with Lyapunov stability theory for linear time-invariant systems. Specifically, the system is asymptotically stable at the steady state—meaning all perturbations decay to zero over time with no parametric oscillation or divergence—if and only if all eigenvalues $\lambda_i$ ($i=1,2,...,6$) of the dynamical matrix $\boldsymbol{M}$ have strictly negative real parts, i.e., $\text{Re}(\lambda_i) < 0$ for all $i$. This criterion ensures that any tiny perturbation of the steady state decays exponentially with time, and the slow-to-fast light conversion effects demonstrated in this work can be stably observed in experiments. Although the analytical stability condition can be derived via the Routh-Hurwitz theorem [2], the resulting closed-form expression is overly cumbersome to present here. We therefore identify the system's stable and unstable parameter regimes through numerical calculation of the eigenvalues of the dynamical matrix $\boldsymbol{M}$.

\section{Numerical Results}
This section presents numerical simulations of the system’s optical responses, focusing on the effects of magnon introduction and parameter tuning on Fano resonances and slow-to-fast light conversion. Unless explicitly specified otherwise in subsequent calculations, the parameter values are set as follows: \(L = 10 \ \mathrm{mm}\), \(\lambda = 1064 \ \mathrm{nm}\) (cavity mode wavelength), \(m = 10 \ \mathrm{m g}\), \(r = 10 \ \mathrm{\mu m}\), \(l = 200\), \(\omega_\phi/2\pi = 0.5 \ \mathrm{MHz}\), \(\omega_m = 1.15\omega_\phi\), \(g_{m} = 1.15g_\phi\), \(\Delta_a = 0.9\omega_\phi\), \(\kappa_a /2\pi= 1.5\times10^5 \mathrm{Hz}\), \(\kappa_\phi/2\pi = \kappa_m/2\pi = 3.5\times 100 \ \mathrm{Hz}\), \(P_{c} = 10 \ \mathrm{mW}\). With the above parameter values, we can calculate that \(g_{m}/2\pi = 0.28\ \mathrm{Hz}\), which is of the same order of magnitude as theoretical calculations and experimental values \cite{PRL2016Cavity}; the finesse of the cavity is \(F_a = \pi c/(\kappa_a L) \approx 1 \times 10^5\); the finesse of the magnon is \(F_m = \omega_m/\kappa_m \approx 1.64 \times 10^3\).

\begin{figure}[htbp!]
\centering
\includegraphics[width=0.5\textwidth]{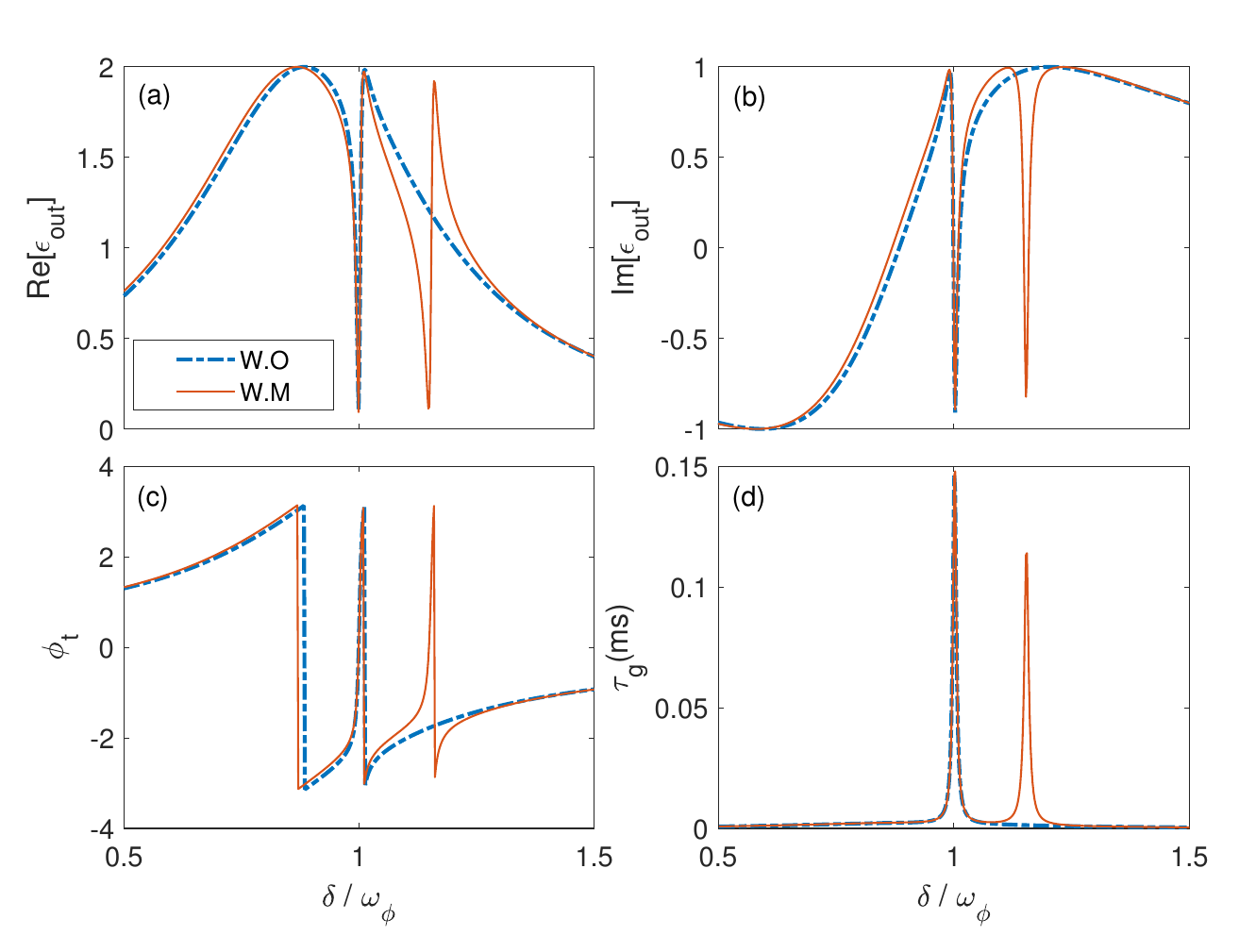}
\caption{Comparison of the optorotational system with magnons (red solid line) and without magnons (blue dashed line). Panels (a)–(d) show the influence of magnon introduction on the absorption (\(\text{Re}[\varepsilon_{\text{out}}]\)), dispersion (\(\text{Im}[\varepsilon_{\text{out}}]\)), phase (\(\phi_t = \arg(t_p)\)), and group delay (\(\tau_g\)) of the probe field, respectively.}
\label{basic1}
\end{figure}
To validate the fundamental impact of integrating magnons into the optorotational system, we first compare the optical responses of the system with and without magnons, focusing on optical parameters that directly reflect light-matter interaction dynamics: absorption, dispersion, phase, and group delay. This comparison serves as the foundation for verifying the effectiveness of magnon integration and exploring subsequent light speed control mechanisms. Fig. \ref{basic1}(a) presents the absorption of the probe field, characterized by \(\text{Re}[\varepsilon_{\text{out}}]\). Two distinct minima are observed in the absorption spectrum after adding magnons, corresponding to normalized probe detunings \(\delta/\omega_\phi = 1.0\) and \(1.15\)—these detunings match the mechanical rotational frequency and magnon frequency, respectively. The minima are associated with OMIT and optomagnetically induced transparency, respectively. Additionally, asymmetric Fano resonances emerge when the effective detuning deviates significantly from these rotational or magnon frequencies \cite{PRA2013Fano}, a result of quantum interference between narrow transparency windows and broadband cavity modes. Fig. \ref{basic1}(b) and (c) display the dispersion (\(\text{Im}[\varepsilon_{\text{out}}]\)) and phase (\(\phi_t = \arg(t_p)\)) of the probe field, respectively. The dispersion and phase profiles near the magnon frequency exhibit features analogous to those induced by optorotational coupling, confirming the effective integration of magnons into the system. Notably, the sharp phase variation observed in Fig. \ref{basic1}(c) gives rise to a significant group delay in Fig. \ref{basic1}(d) \cite{LPR2010Slow}, which is a direct consequence of strong phase dispersion in the vicinity of the transparency windows. The significant group delay induced by phase dispersion demonstrates that the integrated system retains the light speed control capability of traditional optomechanical systems while gaining new tunable degrees of freedom.
\cite{PRB2016Optomagnonics}.
\begin{figure}[htbp!]
    \centering
    \includegraphics[width=0.5\textwidth]{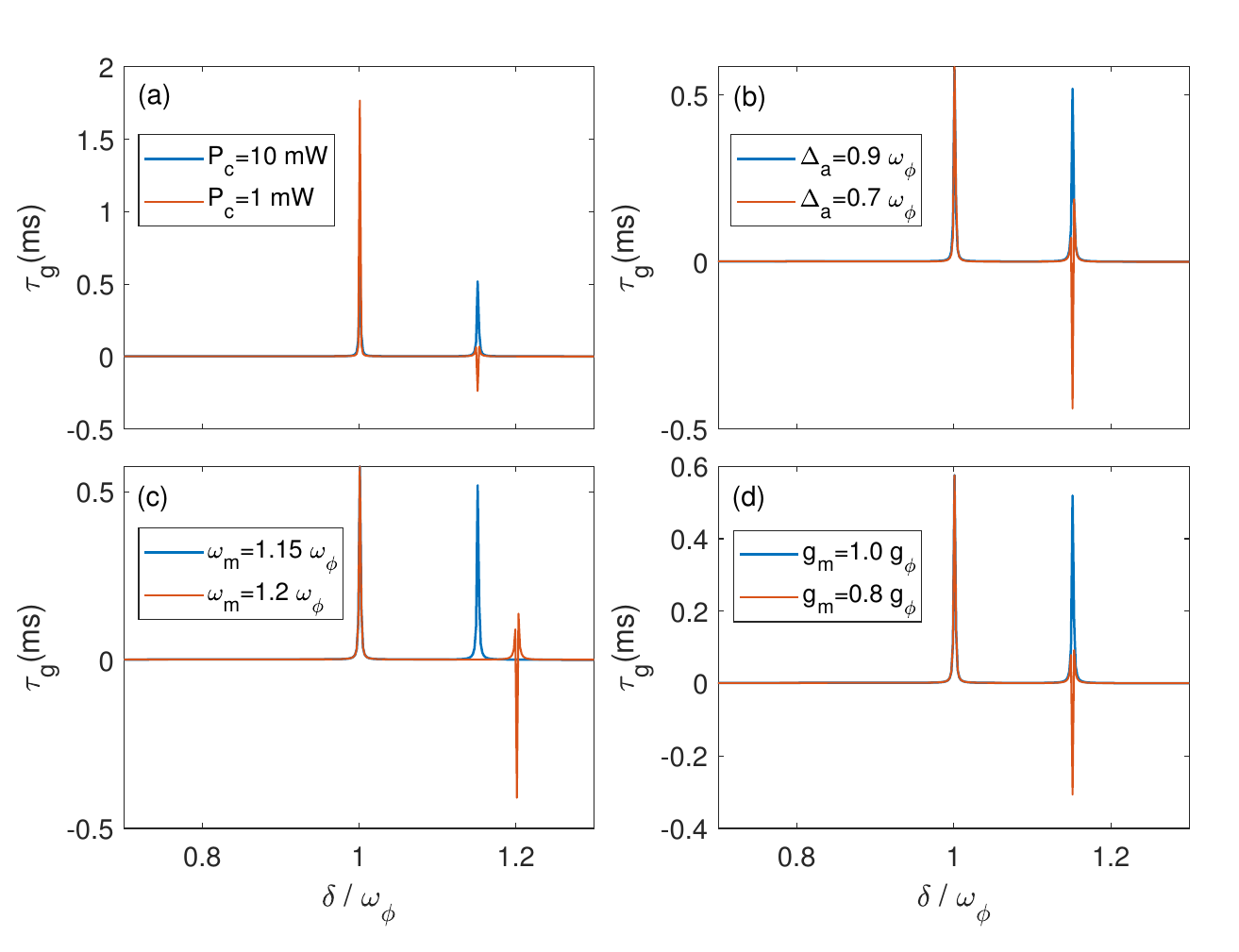}
    \caption{Slow-to-fast light conversion diagrams. Conversion between slow and fast light is achieved by tuning the intensity (a) and frequency (b) of the control field, the magnetic field strength (c), and the optomagnonic coupling (d).}
    \label{basic2}
\end{figure}

Building on the verified effectiveness of magnon integration, we next explore key tunable parameters for slow-to-fast light conversion. To fully exploit the system’s controllability, we select four representative regulatory dimensions—control field intensity, control field frequency, external magnetic field strength, and optomagnonic coupling strength—that cover both optical and magnetic control channels. This design aims to validate the system’s flexibility and address the tunability limitations of traditional optomechanical systems. First, Fig. \ref{basic2}(a) and (b) realize the conversion by tuning the intensity and frequency of the control field, respectively. In Fig. \ref{basic2}(a), we set the control field power to \(P_c=10 \ \mathrm{mW}\) and \(P_c=1 \ \mathrm{mW}\) separately, and it can be found that the group delay associated to magnon changes from the maximum \(\tau_g=1.15 \ \mathrm{ms}\) to the minimum \(\tau_g=-0.24 \ \mathrm{ms}\), which means that we have achieved slow-to-fast light conversion by adjusting the control field power. On the other hand, in Fig. \ref{basic2}(b), we can also realize the slow-to-fast light conversion of the group delay associated to magnon from the maximum \(\tau_g=1.15 \ \mathrm{ms}\) to the minimum \(\tau_g=-0.44 \ \mathrm{ms}\) by changing cavity detuning from the control field from \(\Delta_a=0.9\omega_\phi\) to \(\Delta_a=0.7\omega_\phi\). In addition, Fig. \ref{basic2}(c) enables slow-to-fast light conversion at different probe frequencies through adjusting the magnetic field strength applied to the YIG sphere; specifically, by varying the magnetic field strength, we tune the magnon frequency from \(\omega_m=1.15\omega_\phi\) to \(\omega_m=1.2\omega_\phi\), and observe that the group delay associated to magnon changes from the maximum \(\tau_g=1.15 \ \mathrm{ms}\) to the minimum \(\tau_g=-0.41 \ \mathrm{ms}\).  Finally, Fig. \ref{basic2}(d) demonstrates that slow-to-fast light conversion can be directly achieved by adjusting \(g_m\)—the optomagnonic coupling strength; specifically, by tuning \(g_m\) from \(g_m=1 g_\phi\) to \(g_m= 0.8 g_\phi\), we realize the change of the group delay associated to magnon from the maximum \(\tau_g=1.15 \ \mathrm{ms}\) to the minimum \(\tau_g=-0.31 \ \mathrm{ms}\). This adjustment of \(g_m\) can be realized via experimental means such as tuning the relative position between the YIG sphere and the optical cavity \cite{NC2025Unidirectional} or adjusting the direction of the external bias magnetic field.

These results highlight the system’s superior tunability: control field intensity/frequency tuning offers straightforward optical control paths compatible with existing optomechanical setups; magnetic field tuning leverages the intrinsic frequency sensitivity of magnons \cite{PRA2020Tunable}, enabling non-contact and wide-range frequency shifting without modifying mechanical structures. This multi-channel control capability significantly outperforms pure optomechanical systems limited to mechanical or cavity detuning control.
\begin{figure*}[htbp!]
\centering
\includegraphics[width=0.8\textwidth]{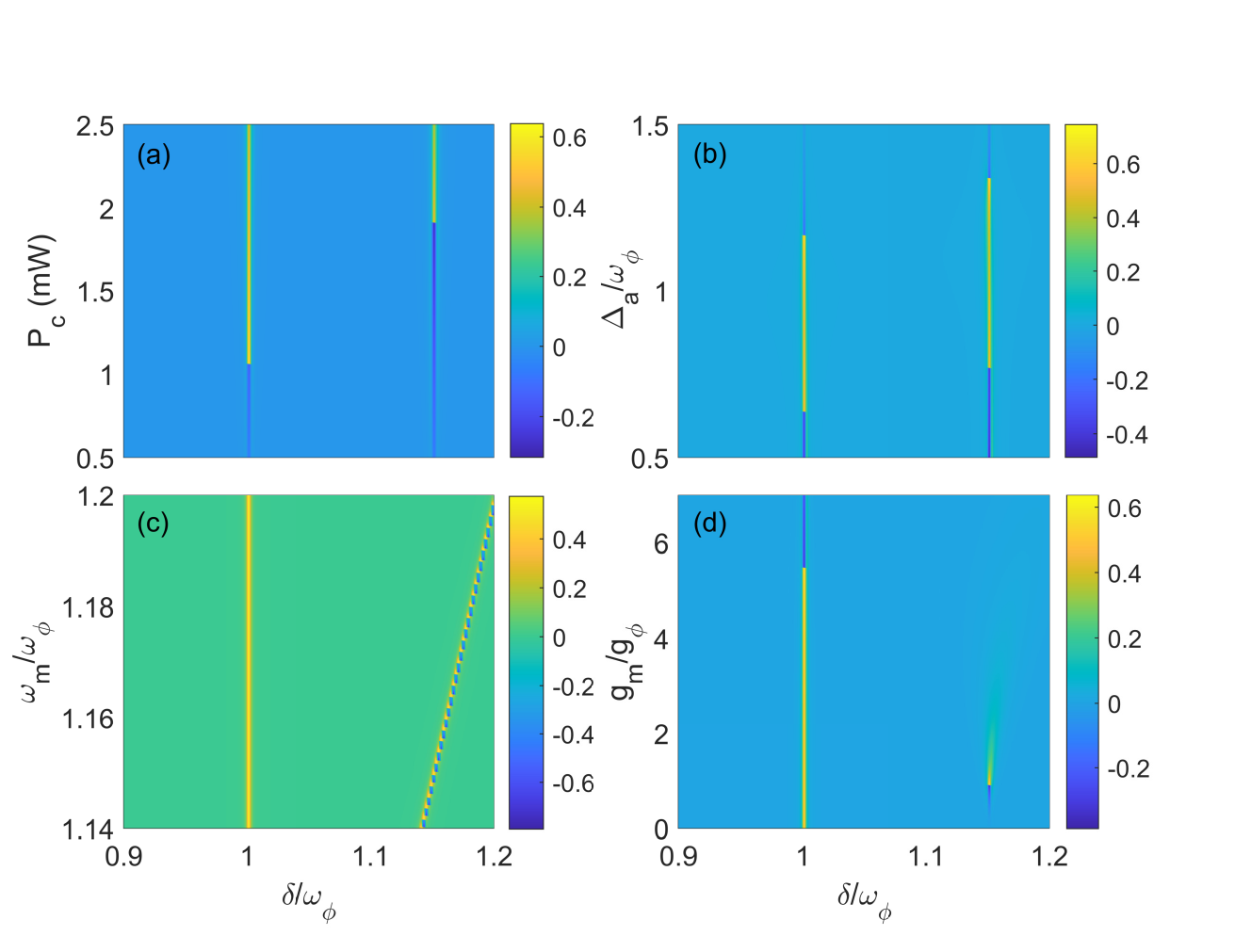}
\caption{Slow-to-fast light conversion via continuous parameter tuning. Conversion between slow and fast light is achieved by continuous adjustment of the control field intensity and frequency, the magnetic field strength, and the optomagnonic coupling. The values of the colorbar denote the group delay \(\tau_g\) (ms).}
\label{basic3}
\end{figure*}

Parameter tuning (Fig. \ref{basic2}) validates the feasibility of slow-to-fast light conversion, while continuous scanning of the four key parameters captures its dynamic evolution, quantifies the control range, and visualizes the group delay to enable explicit identification of conversion thresholds and stable control regions. For detailed insight into parameter influences, Fig. \ref{basic3} illustrates slow-to-fast light switching under their continuous variations. Tuning the control field power (Fig. \ref{basic3}(a)) enables slow-to-fast light conversion at both the magnon frequency and the rotational frequency. Specifically, a slow-to-fast conversion point for the group delay associated to rotation can be observed at a control field power of \(P_c=1 \ \mathrm{mW}\), while the corresponding conversion point for the group delay associated to magnon emerges at \(P_c=1.15 \ \mathrm{mW}\), highlighting the distinct power thresholds for the two modes.

Tuning the control field frequency (Fig. \ref{basic3}(b)) adjusts its detuning with respect to the cavity mode, enabling conversion at both rotational and magnon frequencies; notably, bidirectional conversion (slow-to-fast and fast-to-slow) can occur both at the rotational frequency and the magnon mode frequency. In detail, the group delay associated to rotation exhibits two slow-to-fast conversion points at detunings of \(\Delta_a=0.64 \omega_\phi\) and \(\Delta_a=1.17 \omega_\phi\) between the control field and cavity mode, whereas the group delay associated to magnon presents two conversion points at \(\Delta_a=0.78 \omega_\phi\) and \(\Delta_a=1.34\omega_\phi\), verifying bidirectional tunability across both frequency domains.

Increasing the magnon frequency (Fig. \ref{basic3}(c)) leads to a linear shift in the corresponding probe frequency for light speed switching. When the magnon frequency is continuously varied from \(\omega_m=1.14\omega_\phi\) to \(\omega_m=1.2\omega_\phi\), the group delay associated to rotation remains unchanged, while the group delay associated to magnon shows alternating positive and negative values. This allows us to achieve multiple slow-to-fast light switching points within this frequency range. Meanwhile, this indicates that higher precision is required to control the group delay associated to rotation by tuning the magnon frequency.
\begin{figure*}[htbp!]
\centering
\includegraphics[width=0.8\textwidth]{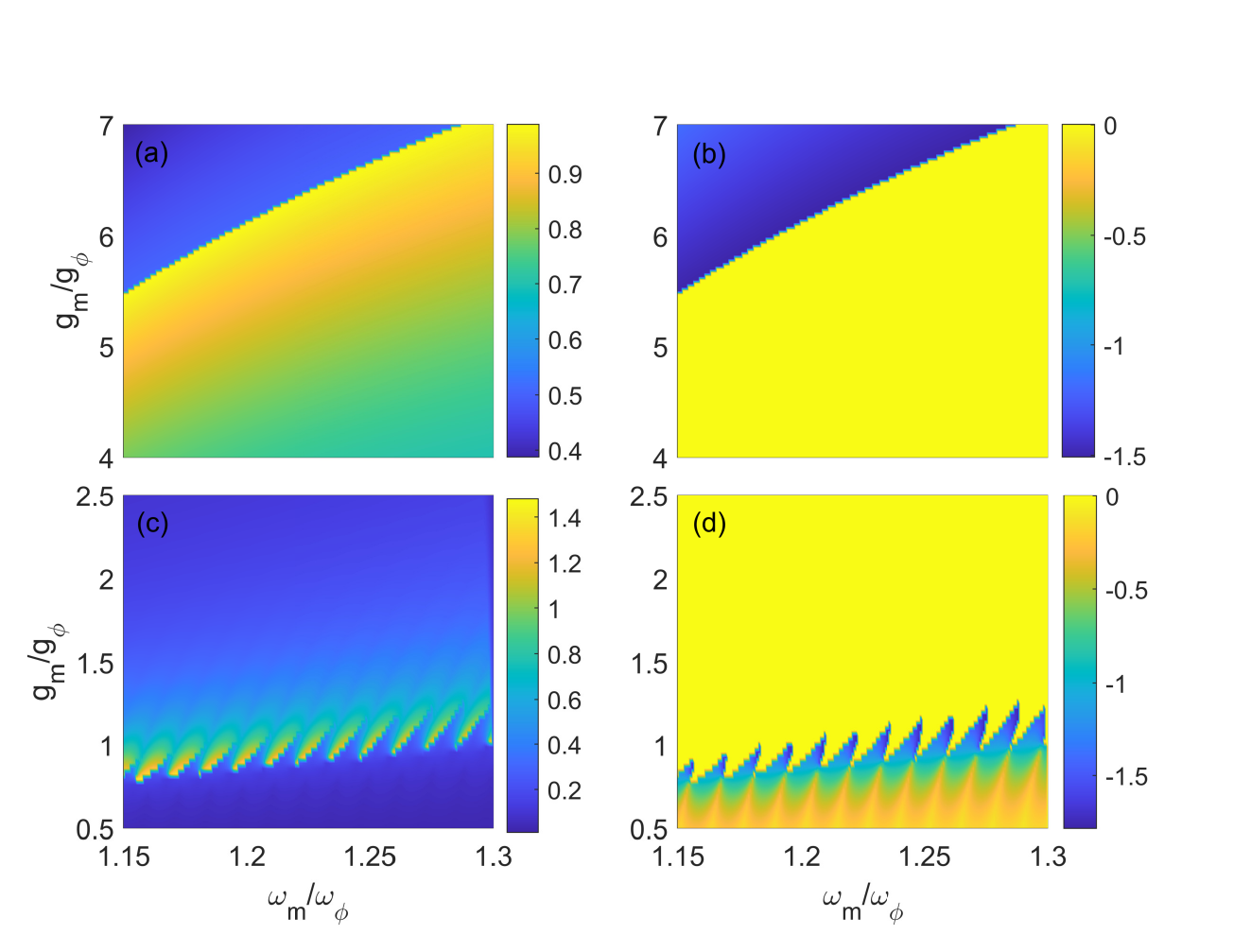}
\caption{(a) and (b) denote the maximum and minimum group delays associated with the rotational mode; (c) and (d) denote those associated with the magnon mode. For each pair of optomagnonic coupling strength \(g_m\) and magnon frequency \(\omega_m\), the maximum and minimum group delays (\(\tau_g\)) presented in the figures are obtained by continuously sweeping the detuning \(\delta\) spanning the rotational and magnon frequencies, followed by extracting the extreme values of \(\tau_g\) within each respective frequency regime. The values of the colorbar denote the group delay \(\tau_g\) (ms).}
\label{basic4}
\end{figure*}

Modulating the optomagnonic coupling (Fig. \ref{basic3}(d)) enables slow-to-fast light conversion at both the rotational and magnon frequencies.  A slow-to-fast light conversion point, associated with rotation, is observed at an optomagnonic coupling strength of \(g_m=5.5 g_\phi\). This is valuable for regulating hard-to-control mechanical degrees of freedom. Separately, the corresponding conversion point for the group delay associated to magnon appears at \(g_m=0.9 g_\phi\). Beyond the conversion threshold, we observe distinct and notable evolutions  specifically in the group delay associated to magnon  with increasing \(g_m\): its magnitude decreases gradually, and its distribution deviates from a linear profile to take on a divergent form. Given that an increase in optomagnonic coupling strength induces the broadening of the magnon-mode effective energy level linewidth via nonlinear dynamics \cite{PRA2016Coupled}, the width and magnitude of the group delay can thus reflect certain information about the energy level width and the coupling strength of the system, offering a means to probe spectral broadening in the optomagnonic subsystem.

Investigation of continuous parameter variation yields key dynamic insights: first, the results of control-field power tuning confirms optomagnonic coupling as an independent control channel, avoiding crosstalk with rotational modes; second, tuning of magnetic parameters offers greater flexibility than the resonant coupling of mechanical modes.

To further quantify the performance of slow-to-fast light conversion and compare the stability of different modes, we extract the maximum and minimum group delays for rotational and magnon modes. We focuses on the extremal values of light speed control, which are critical for evaluating the system’s application potential in scenarios such as optical buffering (requiring large positive group delays) and signal advancement (requiring large negative group delays). We plot the phase diagrams of the maximum and minimum group delays using the magnon frequency and optomagnonic coupling strength as parameters, as shown in Fig. \ref{basic4}. From Fig. \ref{basic4}(a) and \ref{basic4}(b), we observe a distinct boundary line separating the dark blue and yellow regions, which is exactly the group delay boundary line; this boundary exhibits an approximate linear sawtooth-like relationship: larger positive group delays can be achieved below this boundary line (corresponding to the yellow region in Fig. \ref{basic4}(a)), while larger negative group delays are obtainable above this boundary line (corresponding to the dark blue region in Fig. \ref{basic4}(b)). Thus, this relatively stable boundary line delineates two equally stable regions and serves as the threshold for slow-to-fast light conversion regulated by magnetic parameters. In contrast, while the group delay curves for the magnon mode (Figs. \ref{basic4}(c) and (d)) also enable slow-to-fast light conversion, the boundary in these curves is rather complex and less stable. This further verifies that regulating the magnon-associated group velocity via magnetic parameters requires higher precision, whereas controlling the rotational-mode-associated group velocity through magnetic parameters is more stable. This further indicates that magnetic parameters can stably regulate rotational-mode light speed conversion—an advantage conducive to practical applications.

\section{Conclusions}
This work theoretically proposes a hybrid optomagnonic-Laguerre-Gaussian rotational system, integrating cavity photons, mechanical rotation, and YIG magnons to break the fixed-frequency tunability limit of traditional optomechanical slow-to-fast light conversion. Slow-to-fast light conversion is dynamically tunable via four parameters (control field power, control field frequency, external magnetic field, and optomagnonic coupling strength): the external magnetic field enables linear tuning of the magnon-mode conversion frequency, while optomagnonic coupling stabilizes rotational-mode conversion, addressing the tunability limitations of optomechanics.  Notably, we can control the rotation-associated group delay solely by tuning magnetic parameters. Further numerical calculations reveal that stable group delay boundaries exist at the rotational frequency, making magnons ideal for mediating mechanical dynamics, whereas higher regulation precision is required at the magnon-mode frequency.  Experimental feasibility is supported by mature technologies: high-topological-charge LG beams can be generated via spiral phase plates \cite{JO2013Generation}, and YIG spheres host low-damping magnon modes \cite{PRRSPL2022Cavity}. This work provides a flexible platform for multi-frequency light speed control, with potential applications in all-optical buffering \cite{PRA2022Alloptical}, quantum communications \cite{PRXQ2021Quantum}, and high-sensitivity sensing \cite{OE2019Magnetically}.
PRL2010Strong,
\begin{acknowledgments}
This work was supported by National Natural Science Foundation of China (Grant No. 12104296).
\end{acknowledgments}

\section{References}

\bibliography{magnonEnglish}

\end{document}